\documentclass [aps, prl, twocolumn, superscriptaddress]{revtex4}
\usepackage{graphicx}
\usepackage{amsmath, bbm,amssymb}
\begin{document}

\newcommand{\bra}[1]{\langle #1|}
\newcommand{\ket}[1]{|#1\rangle}
\newcommand{\braket}[2]{\langle#1|#2\rangle}

\title{Optimal control of a qubit coupled to
a non-Markovian environment}
\author{P. Rebentrost}\email{rebentr@fas.harvard.edu}
\affiliation{Department of Chemistry and Chemical Biology, Harvard University, 12 Oxford St., Cambridge, MA 02138, USA}
\affiliation{Department
Physik, ASC and CeNS, Ludwig-Maximilians-Universit\"at, Theresienstr.  37,
80333 M\"unchen, Germany}
\affiliation{IQC and Department of Physics
and Astronomy, University of Waterloo, 200 University Ave W, Waterloo,
ON, N2L 3G1, Canada}
\author{I. Serban}
\affiliation{Department
Physik, ASC and CeNS, Ludwig-Maximilians-Universit\"at, Theresienstr.  37,
80333 M\"unchen, Germany}
\affiliation{IQC and Department of Physics
and Astronomy, University of Waterloo, 200 University Ave W, Waterloo,
ON, N2L 3G1, Canada}
\author{T. Schulte-Herbr\"uggen}
\affiliation{
Department of Chemistry, Technische Universit\"at M\"unchen,
Lichtenbergstrasse 4, 85747 Garching, Germany}
\author{F.K. Wilhelm} \email{fwilhelm@iqc.ca}
\affiliation{IQC and Department of Physics and Astronomy, University
of Waterloo, 200 University Ave W, Waterloo, ON, N2L 3G1, Canada}
\date{\today}
\begin{abstract}
A central challenge for implementing quantum
computing in the solid state is decoupling the qubits from the
intrinsic noise of the material. We  investigate the implementation of quantum gates
for a paradigmatic, non-Markovian model: A single qubit coupled to a
two-level system that is exposed to a heat bath. We systematically search
for optimal pulses using a generalization of the novel open systems
Gradient Ascent Pulse Engineering (GRAPE) algorithm. We show and
explain that next to the known optimal bias point of this model, there
are optimal shapes which refocus unwanted terms in the Hamiltonian.
We study the limitations of controls set by the decoherence properties.
This  can lead to a significant improvement of quantum operations in hostile
environments.
\end{abstract}
\maketitle

A promising class of candidates for the practical realization of scalable quantum
computers are solid state quantum devices based on superconductors
\cite{Bertet05b,Astafiev04,Simmonds04,Wallraff04,Vion02} and lateral
quantum dots \cite{Hayashi03}. A key challenge to overcome in this
enterprise is the decoherence induced by the coupling to the macroscopic bath of
degrees of freedom not used for quantum computation
(see e.g.~ref.~[\onlinecite{Nato06II}] for a recent review).
Many of these decoherence sources can be engineered at the origin.
In the case of intrinsic slow noise originating from two-level
fluctuators (TLFs) this is much harder \cite{Jung04,Zorin96}, albeit
not impossible \cite{Eroms06,Steffen06}. Thus, in order to
advance the limitations of coherent quantum manipulations in the solid
state, it is imperative to find strategies which accomodate this kind
of noise. A number of methods have been proposed by intuition and
analogies to different areas, such as dynamical decoupling
\cite{Faoro04,PRAR05}, the optimum working point strategy
\cite{Bertet05b,Vion02,Ithier05}, and NMR-like approaches
\cite{Collin04}. Even in light of their success, it is by no means
clear, whether even better strategies can be formulated and, on a more
general level, where the limits of quantum control under hostile
conditions are reached.

We resort to numerical methods of optimal control. The closed systems GRAPE
(gradient ascent pulse engineering) algorithm
\cite{Khaneja05} has been proven useful in spin and pseudo-spin systems \cite{PRA05},
an important example of the latter being coupled Josephson devices \cite{Spoerl05}.
It was recently extended to open yet strictly Markovian systems \cite{Tosh06}.
Other recent optimal control results also include the presence of noise and decoherence.
General pure dephasing was considered in \cite{Gordon08}.
Ref.~\cite{Mottonen06} treated a semiclassical random-telegraph noise
(RTN) model in the high-temperature limit, while ref.~[\onlinecite{Montangero06}]
focussed on two qubits and classical $1/f$ noise. Optimal state transfer in the spin-boson model was
considered in ref.~\cite{Jirari06}. Ref.~[\onlinecite{Grace06}] optimized qubit gates in the presence of
finite-dimensional, dissipation-free environments.
In this Letter, we generalize the GRAPE algorithm to include
a complex environment leading to non-Markovian qubit dynamics and non-Gaussian noise.
We show that next to an optimal working point there is also an
optimal pulse shape and optimal gate duration. Accelerating the fluctuations can
improve the gate fidelity. We discuss the physics ultimately limiting the gate
performance no longer correctable by pulse shaping.

\emph{Model and method.$-$}In macroscopic samples, $1/f$ noise
\cite{Harlingen04,Jung04,Zorin96} occurs in most observables. The Dutta-Horn model
\cite{Dutta81,Weissman88} explains this phenomenon by the (classical)
superposition of TLFs which randomly jump
between their states,  a process known as random telegraph noise
(RTN).  In small, clean samples the discrete nature of the noise
process from a single dominating fluctuator \cite{Wakai87,Bertet05b}
can be resolved. This leads to semi-phenomenological Hamiltonians
\cite{Paladino02,PRL05,Grishin05}.

We specifically model a qubit coupled to a single TLF by ${H}={H}_S+{H}_I +{H}_B$.
${H}_S$ consists of the qubit and the coupled two-state system, i.e.
\begin{eqnarray}
{H}_S&=&E_1(t){\sigma}_{z}+\Delta{\sigma}_{x}+E_2{\tau}_{z}+\Lambda{\sigma}_{z}{\tau}_{z}.
\end{eqnarray}
${\sigma}_i$ and ${\tau}_i$ are the usual
Pauli matrices operating in  qubit and fluctuator Hilbert space
respectively. $E_1(t)$ is time-dependent and  serves as an external
control. The source of decoherence is the coupling of the fluctuator to the
heat bath, which leads to incoherent transitions between the fluctuator eigenstates,
\begin{equation}
{H}_{I}=\sum_i\lambda_i({\tau}^+{b}_i+{\tau}^-{b}_i^{\dagger}),\quad
{H}_{B}=\sum_i\hbar\omega_ib_i^{\dagger}b_i.
\end{equation} We introduce an Ohmic bath spectrum
$J(\omega)=\sum_i\lambda_i^2
\delta(\omega-\omega_i)=\kappa\omega\Theta(\omega-\omega_c)$ containing
the couplings $\lambda_i$, the dimensionless damping
$\kappa$,  and a high-frequency cutoff $\omega_c$ (which we assume to be the largest frequency in the system).

We are mostly interested in the qubit evolution in the limit of slow TLF
flipping. As a first step we describe the dynamics of the larger
qubit $\otimes$ TLF system by a master equation, tracing out the bath along the lines of \cite{Alicki06,Nato06II}.
We arrive, in the motional narrowing regime
$k_BT > \kappa E_2$, at the Bloch-Redfield equation
\begin{eqnarray}
\dot{{\rho}}(t)&=&\frac{1}{i\hbar}[{H}_S,{\rho}(t)]
+[{\tau}^+,{\Sigma}_1^-{\rho}(t)]
+[{\tau}^-,{\Sigma}_0^+{\rho}(t)]\nonumber\\
&&-[{\tau}^-,{\rho}(t){\Sigma}_1^+]-[{\tau}^+,{\rho}(t){\Sigma}_0^-] \label{eq:master}
\end{eqnarray}
with the different rate tensors ($s=0, 1$)
\begin{equation}
{\Sigma}_s^{\pm}=\frac{1}{({i}\hbar)^2}\int_0^{\infty}dt'\int_0^{\infty}d\omega
J(\omega)(n(\omega)+s)e^{\pm i\omega t'}{\tau}^{\pm}(t').
\label{eq:rates}
\end{equation}
Here, $n(\omega)$ is the Bose function.
Note, that the rate tensors explicitly depend on the control $E_1(t)$
due to the interaction representation of the operators ${\tau}^{\pm}$ in eq.~\ref{eq:rates}.
Since tracing out the TLF at this stage would lead to
an intricate non-Markovian master equation, we treat
the qubit-TLF interactions exactly, i.e. the rate tensors act on the
combined qubit-TLF system.

Our model goes far beyond a simple RTN noise model
\cite{PRAR05} and captures the correlations between qubit and TLF
\cite{Paladino02,Grishin05}. Still, it is useful to
introduce the parameters of the RTN which would result for
$\Lambda\rightarrow0$. The TLF flipping rate is $\gamma=2\kappa
E_2\coth(E_2/T)$, the sum of the excitation and relaxation
rate. It enters the two-point noise spectrum of random telegraph noise
\begin{equation} S(\omega)=\int_{-\infty}^\infty dt\ e^{-i\omega t}\langle
{\tau}_z(t){\tau}_z(0)\rangle_{\rm
eq}=\Lambda^2\frac{\gamma}{\omega^2+\gamma^2}.
\label{eq:powerspectrum}
\end{equation} This is the Fourier transform of the interaction
representation of ${\tau}_z$ assuming the bath in equilibrium. In this limit, we can find
relaxation rates $1/T_1=\frac{\Delta^2}{E^2}S(2E)$ and
$1/T_2=1/2T_1+\frac{E_1^2}{E^2}S(0)$ with $E=\sqrt{\Delta^2+E_1^2}$.
It has been shown in \cite{Vion02} that pure dephasing can be suppressed
by keeping $|E_1| \ll \Delta$ during all manipulations, the optimum working point strategy.
Additionally in the slow flipping regime,
 $1/T_1$-relaxation is small since $S(2\Delta)\simeq\Lambda^2\gamma/4\Delta^2$.

We formulate the control approach by rewriting the master equation
(\ref{eq:master}) as $\dot{\rho}(t) =
-\big(i\mathcal{H}(E_1(t))+\Gamma(E_1(t))\big)\rho(t)$ with the
Hamiltonian commutator superoperator
$\mathcal{H}(E_1(t))(\cdot)=[H(E_1(t)), \cdot]$ and  the relaxation
superoperator $\Gamma$, both time-dependent via the control $E_1(t)$. The
formal solution to the master equation is a linear quantum map
operating on a physical initial state according to $\rho(t) = F(t)\rho(0)$.
Thus $F$ itself follows the operator equation of motion
\begin{equation}\label{FTimeEvolution}
\dot{F}=-\left(i\mathcal{H}+\Gamma\right) F
\end{equation}
with initial condition $F(0)=\mathbbm{1}$, as in
ref.~\cite{Tosh06}. Here, multiplication of quantum maps denotes their
concatenation.
The task is to find control amplitudes $E_1(t)$
with $t\in [0,t_g]$, $t_g$ being a fixed final time,
such that the difference $\delta F=F_U- F(t_g)$ between
dissipative time evolution $F(t_g)$ obeying eqn.~\ref{FTimeEvolution}
and a target unitary map $F_U$ is minimized with respect to the
Euclidean distance $||\delta F||_2^2 \equiv {\rm tr}\left \{ \delta F^\dagger \delta F \right \}$.
Clearly, this is the case, when the trace fidelity
\begin{equation}\label{FidelityOpenSystem} \phi=\rm{Re\, tr}
\left \{ F^{\dagger}_U \; F(t_g)\right \}
\end{equation}
is maximal. Note, that in an open system, one cannot expect to
achieve zero distance to a unitary evolution $F_U$~\cite{Tosh06}.
The goal is to come as close as possible.

We find the pulses by a gradient search. We digitize $F(t_g) \approx
F_N\cdots F_j \cdots F_1$, where the interval $[0,t_g]$ is divided into
$N$ slices of duration $\Delta t$. One finds by
eqn.~\ref{FTimeEvolution}
\begin{equation} F_j=e^{ -\Delta t
\big(i\mathcal{H}(E_1(j))+\Gamma(E_1(j))\big) }
\end{equation} with $E_1(j)$ being the control amplitude in the $j^{\rm
th}$ time slice. The gradient of the fidelity can be computed
in closed form as $\frac{\partial \phi}{\partial E_1(j)} = -\rm{Re}\ tr
\left \{ F^{\dagger}_U F_N \cdots F_{j+1}\;\Delta t \frac{\partial \big(
i\mathcal{H}(E_1(j)) + \Gamma(E_1(j)) \big)}{\partial E_1(j)}F_j \cdots
F_1\right \}.$

We aim at optimizing the evolution of the qubit alone. Therefore,
the TLF is traced out at the end of the full time evolution $F(t_g)$
retaining all degrees of freedom in the intermediate steps $F_j$. One has schematically
\begin{equation}
F^R(t_g)[\cdot]= {\rm tr_{\rm TLF} } \left \{ F(t_g)[\cdot\otimes\rho_{\rm TLF}^{\rm eq}] \right \}
\end{equation}
where the resulting map $F^R$ acts on the space of qubit density matrices
alone. Here, we assume standard factorized initial conditions with the fluctuator in
equilibrium, $\rho_{\rm TLF}^{\rm eq}$. We use $F^R$ in fidelity eq.~(\ref{FidelityOpenSystem})
for the optimization.

\emph{Results and their discussion.$-$} In the above model
we focus on optimizing controls for a $Z$-gate, ${\rm exp}(-i\frac{\pi}{2} \sigma_z)$.
This is a paradigmatic case demonstrating the virtues of our technique: (1)
an error rate up to approximately one order of magnitude lower than
the current optimal working point strategies; (2) the obtainable fidelities
reach the $T_1$ limit of the relaxation model; (3) the optimized
controls relate to optimal times via self-refocussing
 effects---thus showing how open systems GRAPE-derived
controls provide physical insight under structured relaxation
models.  Similar findings can be expected beyond one qubit gates.

\begin{figure}
\includegraphics[width=1\columnwidth]{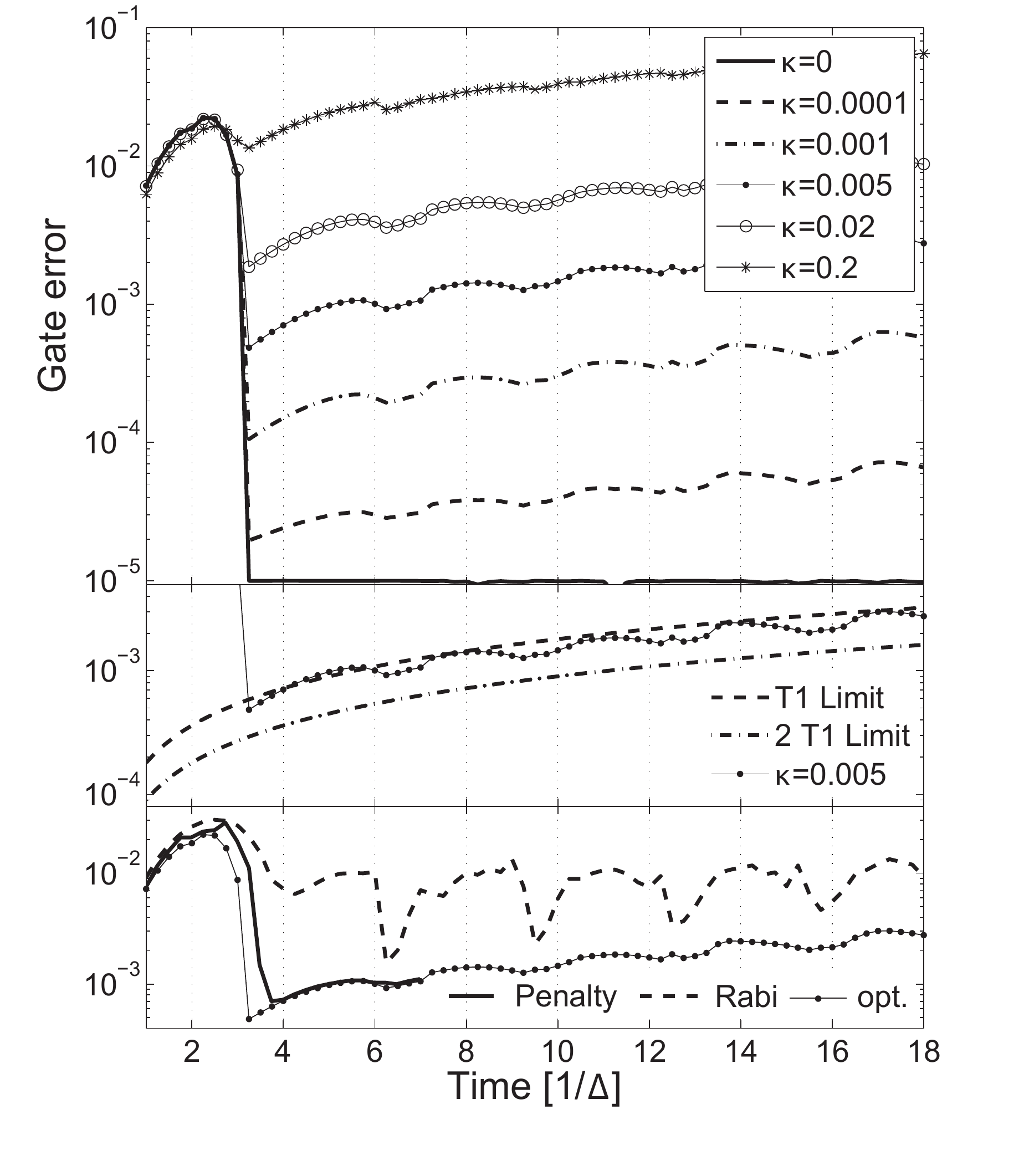}
\caption{Top: Gate error versus pulse time $t_g$ for optimal Z-gate pulses
in the presence of a non-Markovian environment with dissipation strength $\kappa$.
A periodic sequence of minima at around $t_n= n \pi /\Delta$, where $n\ge1$, is obtained.
Middle: The gate error of optimized pulses approaches a limit set by $T_1$ and
$2T_1$, as shown with $\kappa=0.005$.
Bottom: Optimized pulses reduce the
error rate by approximately one order of magnitude compared to  Rabi
pulses for $\kappa= 0.005$. Pulses starting from zero bias and with realistic rise times
(penalty) require only a small additional gate time.
In all figures the system
parameters are  $E_2 = 0.1\Delta$, $\Lambda = 0.1\Delta$ and $T =
0.2\Delta$. \label{fig:Zerror}}
\end{figure}

\begin{figure}
\includegraphics[width=0.49\columnwidth]{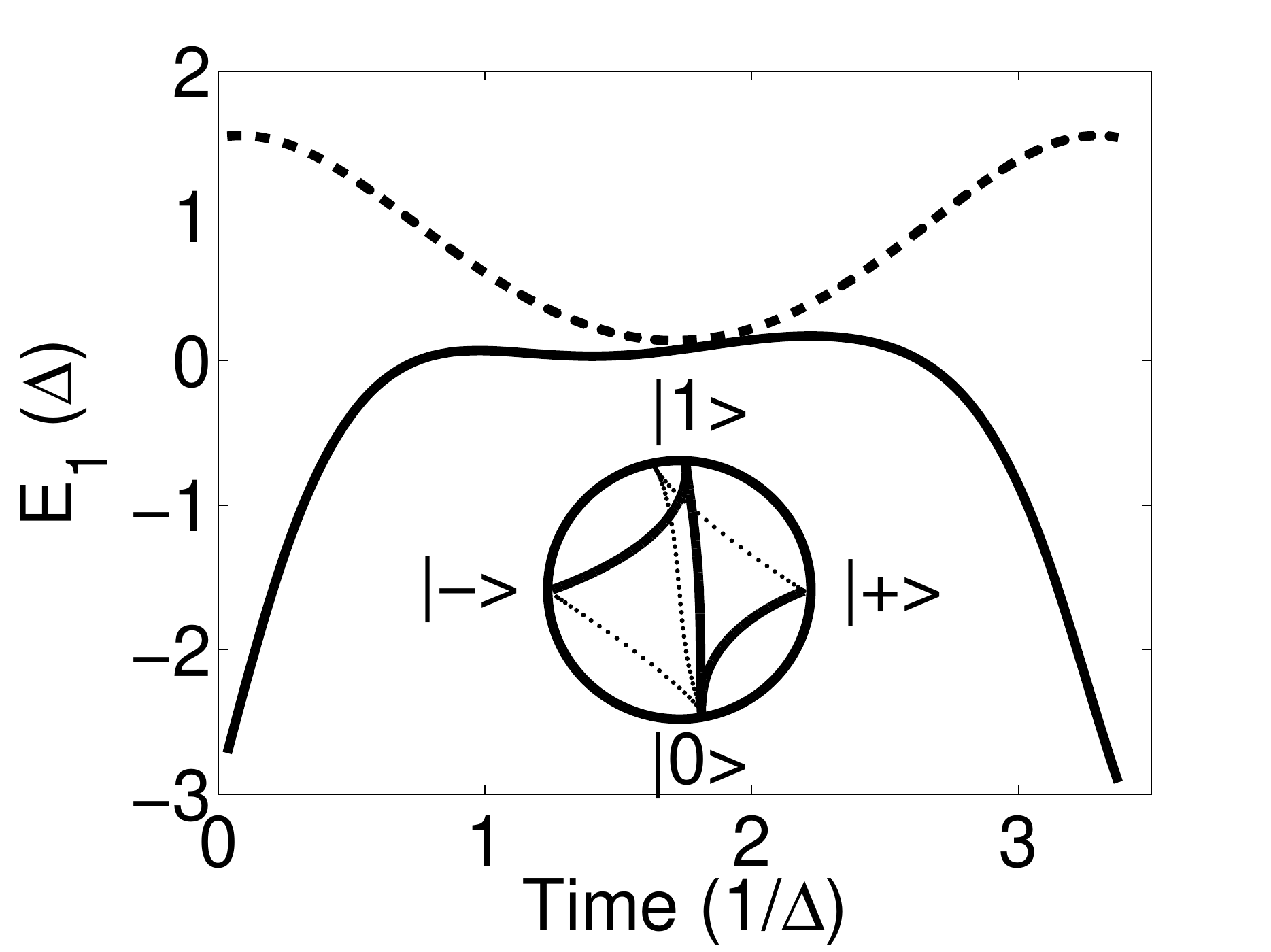}
\includegraphics[width=0.49\columnwidth]{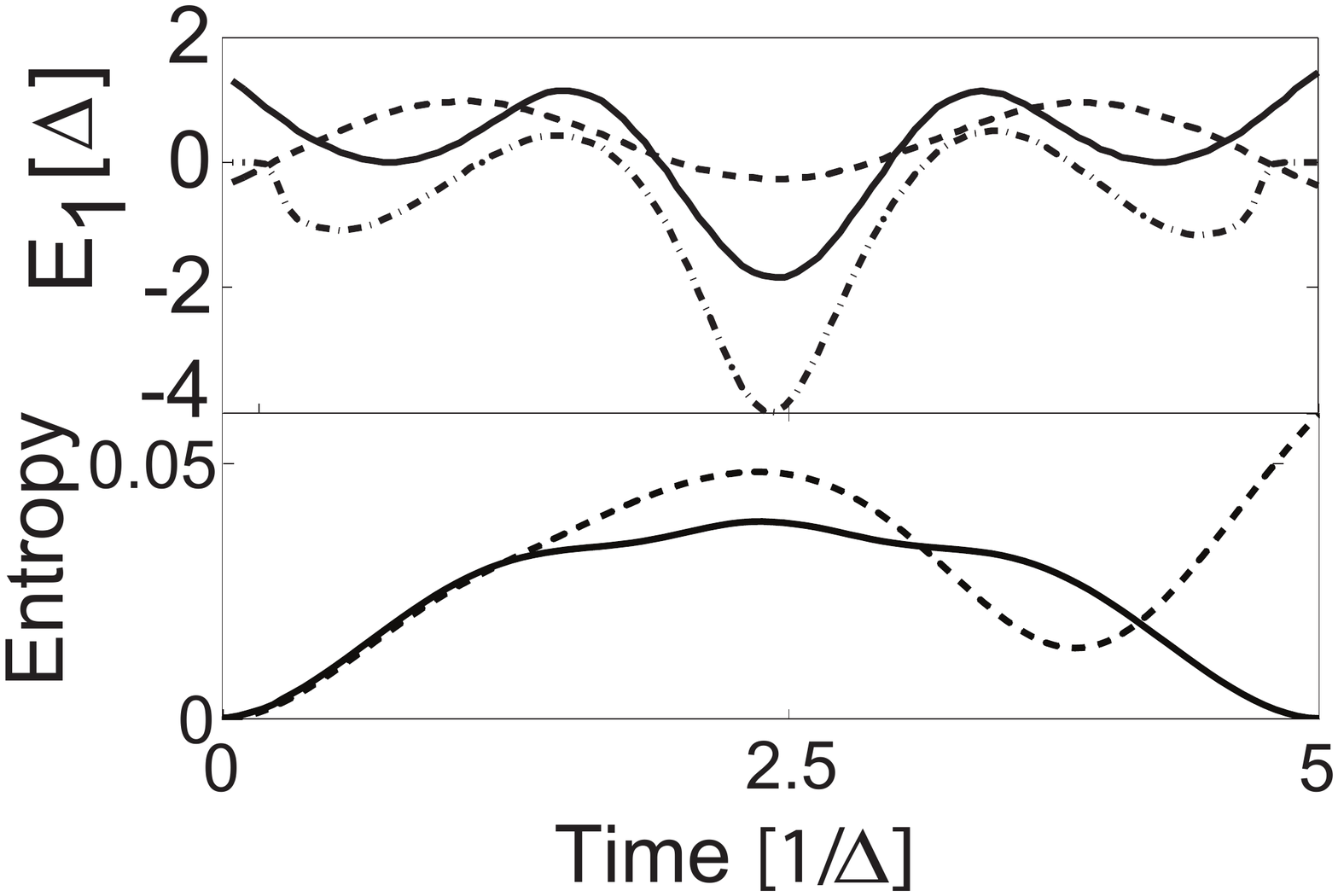}
\caption{
Left: Rabi (- -) and optimized pulse (---) close to the
optimal time at $t_g=3.375/\Delta$ and $\kappa=0.05$. Inset shows the evolution of initial state
 $\rho=|+\rangle\langle +|$, $|\pm\rangle=(|0\rangle\pm|1\rangle)/\sqrt{2}$ on the Bloch sphere
under these pulses.
Right: Comparison of pulse shape and qubit entropy between Rabi (- -), optimized (---),
and penalty-constrained pulse ($-\cdot-$)
at $t_g=5.0/\Delta$  and $\kappa=0$. $E_2=0.1\Delta$, $\Lambda=0.1\Delta$, $T=0.2\Delta$
in all panels. \label{fig:Pulses} }
\end{figure}

An overview of the accessible gate performance as a function of the
duration $t_g$ of the gate is given in fig.~\ref{fig:Zerror} (top).  We
notice that excellent gate performance can be achieved for pulse time
$t_g\gtrsim\pi/\Delta$. This corresponds to the static $\Delta\sigma_x$ inducing
at least a full loop around the Bloch sphere.
Indeed, for the pulse at $t_g=3.375/\Delta$ the evolution consists of a quarter $z$-rotation, a full loop around $x$, and the
second quarter of the $z$-rotation leading to the total half rotation around $z$ necessary
for the $Z$-gate, see Bloch sphere in fig.~\ref{fig:Pulses} (right) for a particular initial state.
At shorter times the pulses cannot use the physical resource provided by the internal evolution
to refocus the qubit.

At longer times the attainable gate performance mildly deteriorates,
depending on the value of $\kappa$.  This indicates that the optimal
pulses are essentially limited by $T_1$ processes at the optimal
working point. We compare the performance to $1-e^{-t_g/T_1}$ with $T_1$
obtained at $E_1=0$. The optimized pulses are able to beat this limit
which indicates that $T_1(E_1=0)$ is a lower bound for the effective
$T_1$,  see fig.~\ref{fig:Zerror}  (middle panel). For clarity, we also
compare to $1/e^{-t/T_{\rm 2, min}}$ with $T_{\rm 2, min} = 2T_1$.

Optimizing the qubit with decoherence
can lead to an improvement of around one order of magnitude over conventional Rabi pulses,
see fig.~\ref{fig:Zerror} (bottom).
The source of error of the Rabi pulses, can be seen as the fast oscillating counter-rotating component in the rotating frame, which is significant for these short, high-amplitude pulses,
which only consist of a few carrier periods.
Only at certain times $t_n\eqsim n\pi/\Delta, n \ge 1$, Rabi pulses perform well.
At these $t_n$ we have n rotations around $x$ due to the static field, refocussing the qubit.
Additional optimization gives moderate results:
in the case $n=2$ the errors are $(1-\phi)_{\rm Rabi}=1.5\cdot 10^{-3}$ and
$(1-\phi)_{\rm GRAPE}=0.9\cdot 10^{-3}$.
At other times, the pulse optimization redirects the $\Delta$ drift to refocus the TLF-field and keeps
the control at the optimal point $E_1=0$ as often as possible.
As a result, the narrow and deep minima of the error for the regular Rabi pulse
become shallow and broad using GRAPE.

Steep rises and offsets of pulses as in fig.~\ref{fig:Pulses} can be smoothened
by adding a penalty function to the fidelity
$\tilde{\phi}=\phi - \int_0^{t_g}\alpha(t) E_1^2(t) dt$.
Although in principle the penalty can be increased with the iterations in the algorithm,
the simple penalty strength
$\alpha(t) = \alpha_0 \left( 2-\tanh\left(\frac{t}{t_0}\right) + \tanh\left(\frac{t_g-t}{t_0}\right) \right)$
was sufficient to avoid offset and short rise times, see fig.~\ref{fig:Pulses} (right).
Here, the overall penalty and the characteristic rise-time parameter are chosen to be $\alpha_0 = 2 \Delta$
and $t_0=0.02/\Delta$, respectively.
Fig.~\ref{fig:Zerror} shows that smooth pulses close to experimentally realistic
settings come at a modest price of $0.5/\Delta$ in additional gate time.
After $t_g\approx3.75/\Delta$ the same gate errors as without the penalty function are obtained.


\begin{figure}
\includegraphics[width=1.0\columnwidth]{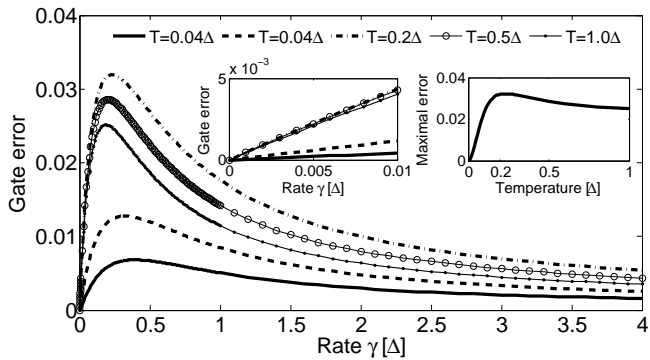}
\caption{Gate error versus TLF rate $\gamma$ for various temperatures for an
optimized pulse with $t_g=5.0/\Delta$.
The left inset is a magnification of the low-$\gamma$ part
of the main plot and reveals the linear behaviour. The right inset
shows the maximum of the curves of the main plot versus temperature.
($E_2=0.1\Delta$ and $\Lambda=0.1\Delta$)
\label{fig:motional}}
\end{figure}

We now analyze the dependence on the bath parameters.  In
fig.~\ref{fig:motional} we can identify a nonmonotonic  dependence of
the error of the optimized pulse on $\gamma$.
At low $\gamma$, we can approximate $(1-\phi)=a+b\gamma$ (left inset).
First, the linear growth with $\gamma$ accounts for the increasing probability
for the TLF to flip at a random time during the evolution. This is
reflected in the power spectrum of the RTN for low $\gamma$, where
essentially $S(\omega\gg\gamma)\simeq \Lambda^2\gamma/\omega^2$,
eq.~\ref{eq:powerspectrum}. Note that  for time-independent $E_1$ and
very long evolution, $T_2$
would be dominated by $S(0)$ in the Markovian limit. However, at very
fast manipulations through an external field, the environment is only
sampled at higher frequencies, $1/t_g \sim \omega\ \gg \gamma$.
Second, it turns out to be that $a\simeq0$.
This reflects that, for a static TLF ($\gamma=0$)
the GRAPE pulse fully compensates for the unknown initial state of the TLF.
At the end of the pulse, the entropy of the fluctuator is completely
pushed out of the qubit, see fig.~\ref{fig:Pulses} (right).

On the other hand, for a high flipping rate $\gamma$, the physics of
motional narrowing sets in. This limits the low frequency noise and
hence the pure dephasing to $S(\omega\ll\gamma)=\Lambda^2/\gamma$ which vanishes for
$\gamma\rightarrow\infty$. Indeed, the high-$\gamma$ part of the error
can be fitted by a law $c+d/\gamma$. The finite limiting value
$c$ captures the residual decoherence which occurs even though the
RTN model, eq.~(\ref{eq:powerspectrum}), suggests absence of noise.
Consequently, there is a $\gamma_{\rm max}$ at which the error is
maximum. Remarkably, $\gamma_{\rm max}\simeq 0.32\Delta \simeq
\Delta/\pi$  independent of any other parameters such as temperature
and pulse length: The attainable performance is worst if the TLF flips
once per free rotation around $x$.

The maximal amplitude of the error $(1-\phi)(\gamma_{max})$
as well as most other fit parameters show a non-monotonic
temperature dependence, see fig.~\ref{fig:motional} (right inset).
It is exponentially suppressed at low $T$ and
saturates to a finite high-temperature limit. The intermediate behaviour
can be related to the equilibrium susceptibility of the TLF,
which is maximum at $T\simeq 2E_2$. The more responsive the TLF
is to perturbations of its $z$-field, the stronger it will influence the qubit.

\emph{Conclusion.$-$}
We have investigated an important model for decoherence in solid-state
systems, a qubit coupled to a two-level fluctuator which itself is
coupled to a heat bath.  Our study is the first to exploit the
explicit dynamics of a complex non-Markovian environment in optimal
control of open systems for implementing quantum gates. The pulses
include offset and rise time limitations of experimental settings.
For a wide range of parameters, we have identified
self-refocusing effects, which are usually only visible at specific optimal
pulse durations but can now be achieved more robustly.
We have shown that both for fast and slow flipping of
the TLF high-fidelity control can be achieved. The full qubit-fluctuator
correlations, embodied in the Hamiltonian, turn out to be a crucial
ingredient.

We gratefully acknowledge support by NSERC discovery grants, by the EU
in  the projects EuroSQIP and QAP as well as by the DFG through SFB
631.


\begin{thebibliography}{10}

\bibitem{Bertet05b}
P. Bertet {\it et al.}, Phys. Rev. Lett. {\bf 95},  257002  (2005).

\bibitem{Astafiev04}
O. Astafiev, Y.A. Pashkin, T. Yamamoto, Y. Nakamura, and J.S. Tsai, Phys. Rev. B
  {\bf 69},  180507(R)  (2004).

\bibitem{Simmonds04}
R.W. Simmonds {\it et al.}, Phys. Rev. Lett.  {\bf 93},  077003  (2004).

\bibitem{Wallraff04}
A. Wallraff {\it et al.}, Nature {\bf 431},  162  (2004).

\bibitem{Vion02}
D. Vion {\it et al.}, Science {\bf 296},  286  (2002).

\bibitem{Hayashi03}
T. Hayashi, T. Fujisawa, H.D Cheong, Y.H. Jeong, and Y. Hirayama, Phys. Rev. Lett.
  {\bf 91},  226804  (2003).

\bibitem{Nato06II}
F. Wilhelm {\it et al.} in Michael E. Flatté and Ionel Tifrea (eds.),
Manipulating Quantum Coherence in Solid State Systems (NATO ASI Series), Springer, Dordrecht, 2007.

\bibitem{Jung04}
S. Jung {\it et al.}, Appl. Phys. Lett. {\bf 85},  768
  (2004).

\bibitem{Zorin96}
A.B. Zorin {\it et al.}, Phys. Rev. B {\bf 53},  13682  (1996).

\bibitem{Eroms06}
J. Eroms {\it et al.}, Appl. Phys. Lett. {\bf
  89},  122516  (2006).

\bibitem{Steffen06}
M. Steffen {\it et al.}, Phys. Rev. Lett. {\bf 97},
  050502  (2006).

\bibitem{Faoro04}
L. Faoro and L. Viola, Phys. Rev. Lett. {\bf 92},  117905  (2004).

\bibitem{PRAR05}
H. Gutmann, F.K. Wilhelm, W.M. Kaminsky, and S. Lloyd, Phys. Rev. A {\bf 71},
  020302(R)  (2005).

\bibitem{Ithier05}
G. Ithier {\it et al.}, Phys. Rev. B {\bf 72}, 134519  (2005).

\bibitem{Collin04}
E. Collin {\it et al.}, Phys. Rev. Lett. {\bf 93},  157005  (2004).

\bibitem{Khaneja05}
N. Khaneja {\it et al.}, J. Magn. Res. {\bf 172},  296  (2005).

\bibitem{PRA05}
T.~Schulte-Herbr{\"u}ggen and A.~K.~Sp{\"o}rl and N.~Khaneja and
S.~J.~Glaser, Phys. Rev. A \textbf{72}, 042331 (2005).

\bibitem{Spoerl05}
A. Sp\"orl {\it et al.}, Phys. Rev. A {\bf 75}, 012302 (2007).

\bibitem{Tosh06}
T. Schulte-Herbr\"uggen {\it et al.},
  quant-ph/0609037 (unpublished).

\bibitem{Gordon08}
G. Gordon, G. Kurizki, and D.A. Lidar, Phys. Rev. Lett. {\bf 101}, 010403 (2008).

\bibitem{Mottonen06}
M. M\"ott\"onen, R. de~Sousa, J. Zhang, and K.B. Whaley, Phys. Rev. A {\bf 73},
022332 (2006); O.-P. Saira, V. Bergholm, T. Ojanen, and M. M\"ott\"onen, Phys. Rev. A {\bf 75},
012308 (2007).

\bibitem{Montangero06}
S. Montangero, T. Calarco, and R. Fazio, Phys. Rev. Lett. \textbf{99}, 170501 (2007).

\bibitem{Jirari06} H. Jirari and W. P\"otz, Phys. Rev. A {\bf 74}, 022306 (2006).

\bibitem{Grace06}
M. Grace {\it et al.}, J. Phys. B: At. Mol. Opt. Phys. \textbf{40}, S103 (2007).

\bibitem{Harlingen04}
D.J. {Van Harlingen} {\it et al.}, Phys. Rev. B {\bf 70},  064517  (2004).

\bibitem{Dutta81}
P. Dutta and P. Horn, Rev. Mod. Phys. {\bf 53},  497  (1981).

\bibitem{Weissman88}
M. Weissman, Rev. Mod. Phys. {\bf 60},  537  (1988).

\bibitem{Wakai87}
R.T. Wakai and D.J. Van Harlingen, Phys. Rev. Lett. {\bf 58},  1687  (1987).

\bibitem{Paladino02}
E. Paladino, L. Faoro, G. Falci, and R. Fazio, Phys. Rev. Lett. {\bf 88},
  228304  (2002).

\bibitem{PRL05}
R. de~Sousa, K.B. Whaley, F.K. Wilhelm, and J. von Delft, Phys. Rev. Lett. {\bf
  95},  247006  (2005).

\bibitem{Grishin05}
A. Grishin, I.~V. Yurkevich, and I.~V. Lerner, Phys. Rev. B {\bf 72},  060509(R)
  (2005); Y.M. Galperin, B.L. Altshuler, J. Bergli, and D.V. Shantsev,
Phys. Rev. Lett. {\bf 96}, 097009 (2006).

\bibitem{Alicki06}
R. Alicki, D.A. Lidar, and P. Zanardi, Phys. Rev. A {\bf 73}, 052311  (2006).

\end{thebibliography}
\end{document}